\documentclass{edm_article}

\usepackage{multirow}
\usepackage{makecell}
\usepackage{graphicx}
\usepackage{booktabs}
\usepackage{amsmath}

\usepackage{algorithm}
\usepackage{algorithmic}
\usepackage{paralist}
\usepackage[english]{babel}
\usepackage{booktabs}

\newtheorem{assumption}{Assumption}

\begin{document}

\title{Identifying At-Risk K-12 Students in Multimodal Online Environments: A Machine Learning Approach}

\numberofauthors{1} 
\author{
\alignauthor
Hang Li, Wenbiao Ding, Zitao Liu\titlenote{The corresponding author.}\\
       \affaddr{TAL Education Group}\\
       \affaddr{16/F, Danling SOHO 6th Danling St}\\
       \affaddr{Haidian District Beijing, China, 100060}\\
       \email{\{lihang4, dingwenbiao, liuzitao\}@100tal.com}
}

\maketitle


\begin{abstract}

With the rapid emergence of K-12 online learning platforms, a new era of education has been opened up. It is crucial to have a dropout warning framework to preemptively identify K-12 students who are at risk of dropping out of the online courses. Prior researchers have focused on predicting dropout in Massive Open Online Courses (MOOCs), which often deliver higher education, i.e., graduate level courses at top institutions. However, few studies have focused on developing a machine learning approach for students in K-12 online courses. In this paper, we develop a machine learning framework to conduct accurate at-risk student identification specialized in K-12 multimodal online environments. Our approach considers both online and offline factors around K-12 students and aims at solving the challenges of (1) multiple modalities, i.e., K-12 online environments involve interactions from different modalities such as video, voice, etc; (2) length variability, i.e., students with different lengths of learning history; (3) time sensitivity, i.e., the dropout likelihood is changing with time; and (4) data imbalance, i.e., only less than 20\% of K-12 students will choose to drop out the class. We conduct a wide range of offline and online experiments to demonstrate the effectiveness of our approach. In our offline experiments, we show that our method improves the dropout prediction performance when compared to state-of-the-art baselines on a real-world educational dataset. In our online experiments, we test our approach on a third-party K-12 online tutoring platform for two months and the results show that more than 70\% of dropout students are detected by the system. 
\end{abstract}

%

\keywords{Dropout, retention, multimodal learning, online tutoring, K-12 education} 

\section{Introduction}
\label{sec:intro}
With the recent development of technologies such as digital video processing and live streaming, there has been a steady increase in the amount of K-12 students studying online courses worldwide. Online classes have become necessary complements to public school education in both developing and developed countries \cite{koller2012moocs,kizilcec2013deconstructing,lykourentzou2009dropout,pereira2019early,li2020siamese,li2020multimodal,10.1145/3366423.3380018}. Different from public schools that focusing on teaching in traditional brick-and-mortar classrooms with 20 to 50 students, online classes open up a new era of education by incorporating more personalized and interactive experience \cite{halawa2014dropout,lee2019machine,rumberger1987high,chen2019multimodal,xu2020automatic}. 

In spite of the advantages of this new learning opportunity, a large group of online K-12 students fail to finish course programs with little supervision either from their parents or teachers. Students drop out of the class may be due to many reasons such as lack of interests or confidence, mismatches between course contents and students' leaning paths or even no immediate grade improvements from their parents' perspectives \cite{lykourentzou2009dropout,mendez2008factors,kim2014understanding}. Therefore, it is crucial to build an early dropout warning system to identify such at-risk online K-12 students and provide timely interventions. 

A large spectrum of approaches have been developed and successfully applied in predicting dropout in Massive Open Online Courses (MOOCs) \cite{kloft2014predicting,sharkey2014process,ramesh2014learning,yang2013turn,balakrishnan2013predicting,borrella2019predict}. However, identifying dropout of K-12 students on online courses are significantly different from MOOCs based attrition prediction. The main differences are summarized as follows:

\begin{itemize}

\item \textbf{watching v.s. interaction}: Even though both learning are conducted in the online environment, learners' engagements on MOOCs and K-12 online platforms vary a lot \cite{halawa2014dropout}. In MOOCs, learners mainly watch the pre-recorded video clips and discuss questions and assignments with teaching assistants on the MOOC forums \cite{gitinabard2018your}. While in K-12 online courses, students frequently interact with the online tutors in a multimodal and immersive learning environment. The tutors may answer students' questions, summarize the knowledge points, take notes for students, etc.  

\item \textbf{spontaneous action v.s. paid service}: Learners on existing popular MOOC platforms such as Coursera\footnote{https://www.coursera.org/}, edX\footnote{https://www.edx.org/}, etc. are adults, who aim at continuing their lifelong learning in higher education and obtaining professional certificates such as Coursera's Specializations and edX's MicroMasters. MOOC learners are typically self-motivated and self-driven. On the contrary, most available K-12 online education choices are commercialized in service industry. Students pay to enroll online tutoring programs to strengthen their in-class knowledge levels and improve their grades in final exams. As a result, there are numerous out-of-class activities involved in K-12 online learning such as follow-ups from personal instructors, satisfaction survey and communications with students' parents, etc. These out-of-class activities rarely appear in MOOC based learning.

\item \textbf{high v.s. low dropout rate}: The dropout rate for MOOC based program is often as high as 70\% - 93\% \cite{koller2012moocs,wang2013exploring} while the dropout rate in K-12 online courses is below 20\%. 
\end{itemize}

Therefore, it is important to study approaches to identify at-risk K-12 online students and build an effective yet practical warning system. However, this task is rather challenging due to the following real-world characteristics:

\begin{itemize}
\item \textbf{multiple modalities}: K-12 online learning is conducted in an immersive and multimodal environment. Students and instructors interact with each other visually and vocally. There are a lot of multimodal factors that may influence the final decisions of dropout, ranging from interaction qualities between students and teachers, teaching speeds, volumes, emotions of the online tutors, etc.

\item \textbf{length variability}: Students join and leave the online platforms independently, which results in a collection of observation sequences with different lengths. A dropout prediction system should be able to (1) make predictions for students with various lengths of learning histories; and (2) handle newly enrolled students.

\item \textbf{data imbalance}: The overall dropout rate for K-12 online classes is usually below 20\%, which makes the training samples particularly imbalanced.
\end{itemize}

The objective of this work is to study and develop models that can be used for accurately identifying at-risk K-12 students in multimodal online environments. More specifically, we are interested in developing models and methods that can predict risk scores (dropout probabilities) given the history of past observations of students. We develop a data augmentation technique to alleviate class imbalance issues when considering the multi-step ahead prediction tasks. We conduct extensive sets of experiments to examine every component of our approach to fully evaluate the dropout prediction performance.

Overall this paper makes the following contributions:

\begin{itemize}
\item We design various types of features to fully capture both in-class multimodal interactions and out-of-class activities. We create a data augmentation strategy to simulate the time-sensitive changes of dropout likelihood in real scenarios and alleviate the data imbalance problem. 
\item We design a set of comprehensive experiments to understand prediction accuracy and performance impact of different components and settings from both qualitative and quantitative perspectives by using a real-world educational dataset.
\item We push our approach into a real production environment to demonstrate the effectiveness of our proposed dropout early warning system.
\end{itemize}

The remainder of the paper is organized as follows: Section \ref{sec:related} discusses the related research work of dropout prediction in both public school settings and MOOCs scenarios. Comparisons with relevant researches are discussed. In Section \ref{sec:problem}, we introduce assumptions when building a practical at-risk student identification system and formulate the prediction task. Section \ref{sec:model}, we describe the details about our prediction framework, which include (1) extracting various types of features from both online classroom recordings and offline activity logs (See Section \ref{sec:features}); and (2) data augmentation technique that helps us create sufficient training pairs and overcomes the class imbalance problem (See Section \ref{sec:label}). In Section \ref{sec:exp}, we (1) quantitatively show that our model supports better dropout predictions than alternative approaches on an educational data derived from a third party K-12 online learning platform and (2) demonstrate the effectiveness of our proposed approach in the a real production environment. We summarize our work and outline potential future extensions in Section \ref{sec:conclusion}.

\section{Related Work}
\label{sec:related}

Dropout prediction and at-risk student identification have been gaining popularity in both the educational research and the AI communities. Understanding the reasons behind dropouts and building early warning systems have attracted a growing interest of academics in the learning analytics area. Broadly speaking, existing research regarding dropout prediction can be categorized by learning scenarios and divided into two categories: (1) public school dropout (See Section \ref{sec:public}); and (2) MOOCs dropout (See Section \ref{sec:mooc}).

\subsection{Public School Dropout}
\label{sec:public}

Education institutions are faced with the challenges of low student retention rates and high number of dropouts \cite{rumberger1987high,lansford2016public}. For examples, in the United States, almost one-third of public high school students fail to graduate from high school each year \cite{monrad2007high,bridgeland2006silent} and over 41\% of undergraduate students at four-year institutions failed to graduate within six years in Fall 2009 \cite{mcfarland2017condition}. Hence, research work has focused on predicting the dropout problem and developing dropout prevention strategies \cite{monrad2007high,pallas1986school,catterall1987social,wood2017predicting,dupere2018high,knowles2015needles,coleman2020better,sullivan2017early}. Zhang and Rangwala develop an at-risk student identification approach based on iterative logistic regression that utilizes all the information from historical data from previous cohorts \cite{zhang2018early}. The state of Wisconsin creates a predictive dropout early warning system for students in grades six through nine and provides predictions on the likelihood of graduation for over 225,000 students \cite{knowles2015needles}. The system utilizes ensemble learning and is built on the steps of searching through candidate models, selecting some subsets of best models, and averaging those models into a single predictive model. Lee and Chung address the class imbalance issue using the synthetic minority over-sampling techniques on 165,715 high school students from the National Education Information System in South Korea \cite{lee2019machine}. Ameri et al. consider different groups of variables such as family background, financial, college enrollment and semester-wise credits and develop a survival analysis framework for early prediction of student dropout using Cox proportional hazards model \cite{ameri2016survival}. 

\subsection{MOOCs Dropout}
\label{sec:mooc}

With the recent boom in educational technologies and resources both in industry and academia, MOOCs have rapidly moved into a place of prominence in the mind of the public and have attracted a lot of research attentions from many communities in different domains. Among all the MOOC related research questions, dropout prediction problem emerges due to the surprisingly high attrition rate \cite{whitehill2017mooc,greene2015predictors,kizilcec2015attrition,jiang2014predicting,rose2014social,xing2016temporal,boyer2015transfer,coleman2015probabilistic,crossley2016combining,halawa2014dropout}. Ramesh et al. treat students' engagement types as latent variables and use probabilistic soft logic to model the complex interactions of students' behavioral, linguistic and social cues \cite{ramesh2014learning}. Sharkey et al. conduct a series of experiments to analyze the effects of different types of features and choices of prediction models \cite{sharkey2014process}. Kim et al. study the in-video dropouts and interaction peaks, which can be explained by five identified student activity patterns \cite{kim2014understanding}. He et al. propose two transfer learning based logistic regression algorithms to balance the prediction accuracy and inter-week smoothness \cite{he2015identifying}. Tang et al. formulate the dropout prediction as a time series forecasting problem and use a recurrent neural network with long short-term memory cells to model the sequential information among features \cite{tang2018time}. Both Yang et al. and Mendez et al. conduct survival analysis to investigate the social and behavioral factors that affect dropout along the way during participating in MOOCs \cite{yang2013turn,mendez2008factors}. Detailed literature surveys on MOOC based dropout prediction are reviewed comprehensively in \cite{taylor2014likely,borrella2019predict}.

In this work, we focus on identifying at-risk students in K-12 online classes, which is significantly distinguished from dropout predictions in either public school or MOOCs based scenarios. In the K-12 multimodal learning environment, the learning paradigm focuses on interactions instead of watching. The interactions come from different modalities, which rarely happen in traditional public schools and MOOC based programs of higher education. Furthermore, as a paid service, K-12 online learning involves both in-class and out-of-class activities and both of them contain multiple factors that could lead to class dropouts. These differences make existing research works inapplicable in K-12 online learning scenarios. To the best of our knowledge, this is the first research that comprehensively studies the dropout prediction problem in K-12 online learning environments from real-world perspectives.

\section{Problem Formulation}
\label{sec:problem}

\subsection{Assumptions}

In order to characterize the K-12 online learning scenarios, we need to carefully consider every cases in the real-world environment and make reasonable assumptions. Without loss of generality, we have the following assumptions in the rest of the paper.

\begin{assumption}[Recency Effect]
\label{as:recency}
Time spans between the date of dropout and the date of last online courses vary a lot. Students may choose to drop the class right after one course or quit after two weeks of no course. Therefore, the per-day likelihood of dropout should be time-aware and the closer to the dropout date, the more accurate the dropout prediction should be.
\end{assumption}

\begin{assumption}[Multi-step Ahead Forecast]
\label{as:multistep}
The real-world dropout prediction framework should be able to flexibly support multi-step ahead predictions, i.e., the next-day and next-week probabilities of dropout.
\end{assumption}

\subsection{The Prediction Problem}
\label{sec:formulation}

In this work, our objective is to predict the value of future status for the \emph{target student} given his or her past learning history, i.e., observations collected from K-12 online platforms. More specifically, let $\mathbf{S}$ be the collection of all students and for each student $s, s \in \mathbf{S}$, we assume that we have observed a sequence of $n_s$ past observation-time pairs $\{ <\mathbf{x}^s_j,t^s_j> \}_{j=1}^{n_s}$, $\mathbf{x}^s_j \in \mathbf{X}^s$, and $t^s_j \in \mathbf{T}^s$, such that $0< t^s_j <t^s_{j+1}$, and $\mathbf{x}^s_j$ is the observation vector made at time ($t^s_j$) for student $s$. $\mathbf{X}^s$ and $\mathbf{T}^s$ represent the collections of observations and timestamps for student $s$. Correspondingly, let $\mathbf{Y}^s$ be the collection of indicators of status (\emph{dropout}, \emph{on-going} or \emph{completion}) of student $s$ at each timestamp, i.e., $\mathbf{Y}^s = \{ y^s_j \}_{j=1}^{n_s}$. Let $\Delta$ be the future time span in multi-step ahead prediction. Time $t^s_{n_s+\Delta}$ ($\Delta > 0$) is the time at which we would like to predict the student's future status $\hat{y}^s_{t^s_{n_s+\Delta}}$. 

Please note that we omit the explicit student index $s$ in the following sections for notational brevity and our approach can be generalized into a large samples of student data without modifications.

\section{The Prediction Framework}
\label{sec:model}
The dropout prediction for K-12 online courses is a time-variant task. A student who just had the class should have a smaller dropout probability compared with a student haven't take any class for two weeks. Therefore, when designing a real applicable approach of dropout prediction,  such recency effect, i.e., Assumption \ref{as:recency}, has to be considered. In this work, we extract both static and time-variant features from different categories to capture the factors leading to dropout events comprehensively (See Section \ref{sec:features}). Furthermore, we create a label augmentation technique that not only alleviates the class imbalance problem when building predictive framework for K-12 online classes, but incorporates the recency effect into label constructions (See Section \ref{sec:label}). The learning of our dropout model is discussed in Section \ref{sec:learning} and the overall learning procedure is summarized in Section \ref{sec:summary}.

\subsection{Features}
\label{sec:features}
In this section, we develop a distinguished set of features for at-risk student identification from the real-world K-12 online learning scenarios, which can be divided into three categories: (1) in-class features that focus on K-12 students' online behaviors during the class (See Section \ref{sec:inclass}); (2) out-of-class features that consider as much as possible real-world factors happened after the class, which may influence the dropout decisions (See Section \ref{sec:post}); and (3) time-variant features that include both historical performance of teachers and aggregated features of student activities within fixed-size windows (See Section \ref{sec:time_variant}).

\subsubsection{In-class Features}
\label{sec:inclass}

Different from adults who continue their learning in higher education on MOOC based platforms, K-12 students come for grade improvements. This intrinsic difference in their learning goals leads to contrasting learning behaviors. Adult learners of MOOCs study independently by various activities, such as viewing lecture videos, posting questions on MOOC forums, etc. This results in various types of in-class click-stream data, which are shown to be effective in dropout prediction in many existing research works \cite{fei2015temporal,taylor2014likely,whitehill2017mooc,boyer2015transfer,crossley2016combining,coleman2015probabilistic,halawa2014dropout}. However, such click based activities barely happen in K-12 online scenarios. Instead, there are frequent voice based interactions between K-12 students and their teachers. The teachers not only make every effort to clarify unsolved questions that students remain from their public schools, but are responsible for arousing students' learning interests and building their studying habits. Therefore, we focus on extracting in-class multimodal features specializing in K-12 tutoring scenarios from the online classroom videos. We categorize our features as follows. Table \ref{tab:example} illustrates some examples of in-class features from different categories.

\begin{itemize}
\item \textbf{Prosodic features}: speech-related features such as signal energy, loudness, Mel-frequency cepstral coefficients (MFCC), etc.

\item \textbf{Linguistic features}: language-related features such as statistics of part-of-speech tags, the number of interregnum words, distribution of length of sentences, voice speed of each sentence, etc.

\item \textbf{Interaction features}: features such as the number of teacher-student Q\&A rounds, the numbers of times teachers remind students to take notes etc.
\end{itemize}

\begin{figure}[!tpbh]
    \centering
    \includegraphics[width=\linewidth] {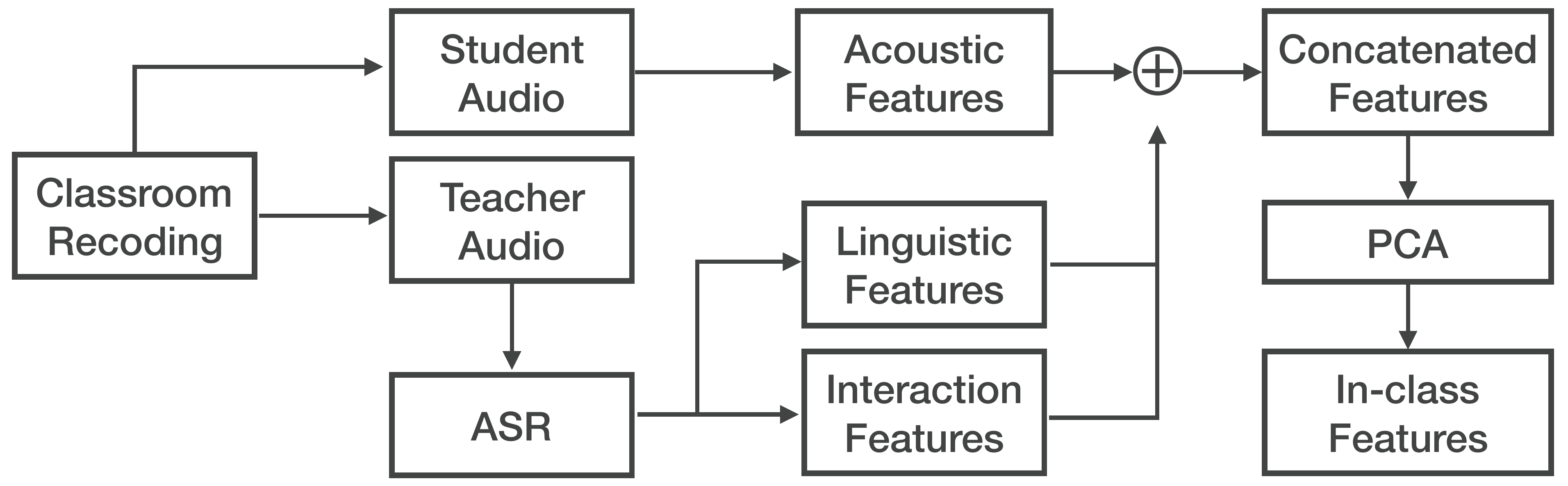}
    \caption{The workflow of our in-class features extraction. ASR is short for automatic speech recognition.}
    \label{fig:inclass_workflow}
\end{figure}

To extract all the features listed in Table \ref{tab:example}, we first extract audio tracks from classroom recordings on both teacher's and student's sides. Then we extract acoustic features directly from classroom audio tracks by utilizing the widely used open-sourced tool, i.e., \emph{OpenSmile}\footnote{https://www.audeering.com/opensmile/}. We obtain classroom transcriptions by passing audio files to a self-trained automatic speech recognition (ASR) module. After that, we extract both linguistic and interaction features from the conversational transcripts. Finally, we concatenate all features from above categories and apply a linear PCA to get the final dense in-class features. The entire in-class feature extraction workflow of our approach is illustrated in Figure \ref{fig:inclass_workflow}.

Please note that due to the benefits of online steaming, both students' and teachers' videos are recorded separately and hence, there is no voice overlap in the video recordings. This avoids the unsolved challenge of speaker diarization \cite{anguera2012speaker}. Similar to Blanchard et al. \cite{blanchard2015study}, we find that publicly available ASR service may yield inferior performance in the noisy and dynamic classroom environments. Therefore, we train our own ASR models on the classroom specific dataset based on a deep feed-forward sequential memory network, proposed by Zhang et al. \cite{zhang2018deep}. Our ASR has a word error rate of 28.08\% in our classroom settings.

\begin{table*}[t]
    \centering
    \caption{List of examples in in-class, out-of-class, and time-variant features.}
    \label{tab:example}
    \begin{tabular}{c|c|l}
    \toprule
    \textbf{Category} & \textbf{Type} & \textbf{Examples} \\ \midrule

    \multirow{15}{*}{In-class} & \multirow{5}{*}{Prosodic} & the average signal energies of student and teacher \\
                               &                           & the average loudness of student and teacher \\
                               &                           & the Mel-frequency cepstral coefficients of audio tracks from student and teacher \\
                               &                           & the zero-crossing rates of student and teacher \\
                               &                           & $\cdots$ \\
    \cline{2-3}
                               & \multirow{5}{*}{Linguistic} & \# of sentences per class of student and teacher  \\
                               &                             & \# of pause words per class of student and teacher \\
                               &                             & average lengths of sentences per class of student and teacher \\
                               &                             & voice speeds (char per second) of student and teacher \\
                               &                             & $\cdots$ \\
    \cline{2-3}
                               & \multirow{5}{*}{Interaction} & \# of teacher-student Q\&A rounds \\
                               &                              & \# of times the teacher reminds the student to take notes and summarization \\
                               &                              & \# of times the teacher asks the student to repeat \\ 
                               &                              & \# of times the teacher clarifies the student's questions \\
                               &                              & $\cdots$ \\
    \midrule
    \midrule

    \multirow{10}{*}{Out-of-class} & \multirow{5}{*}{Pre-class} & \# of days since the student places the online course order \\
                                   &                            & \# of courses in the student's order \\
                                   &                            & \# of conversations between the sales staff and the students (or their parents) \\
                                   &                            & the discount ratio of the student's  order \\
                                   &                            & $\cdots$ \\ 
    \cline{2-3}
                                   & \multirow{5}{*}{Post-class} & \# of follow-ups after the student took the first class \\
                                   &  & \# of words in the latest follow-up report \\
                                   &  & \# of times the student reschedules the class \\
                                   &  & the follow-up ratio, i.e., \# of follow-ups divided by \# of taken courses \\
                                   &  & $\cdots$ \\
    \midrule
    \midrule
    
    \multirow{10}{*}{Time-variant} & \multirow{5}{*}{Historical performance} & \# of courses taught by each individual teacher in total \\
                                  &  & \# of courses the student had in total \\
                                  &  & historical dropout rates \\
                                  &  & historical average time span between classes \\
                                  &  & $\cdots$\\
    \cline{2-3}
                                  & \multirow{5}{*}{Lookback window} & \# of courses taken in past one/two/three weeks\\
                                  &  & \# of courses the student scheduled in past one/two/three weeks \\
                                  &  & \# of positive/negative follow-up reports in past one/two/three weeks \\
                                  &  & the average time span of classes taken in past month \\
                                  &  & $\cdots$\\ 

    \bottomrule
    \bottomrule
    \end{tabular}
\end{table*}

\subsubsection{Out-of-class Features}
\label{sec:post}

As we discussed in Section \ref{sec:intro}, personalized K-12 online tutoring is a paid service in most countries. Besides the course quality itself, there are multiple other factors in such service industry that may change customers' minds to drop the class. Therefore, out-of-class features play an extremely important role in identifying at-risk students in real-world K-12 online scenarios, which are typically ignored in previous literatures. In this work, we collect and summarize all the available out-of-class features and divide them into the following two categories. The illustrative examples are listed in Table \ref{tab:example}.

\begin{itemize}
    \item \textbf{Pre-class features}: Pre-class features capture the students' (or even their parents') behaviors before taking the class, such as purchasing behaviors, promotion negotiations, etc. Examples: the number of rounds of conversation and negotiation before the class, how much the discount student received, etc. 

    \item \textbf{Post-class features}: Post-class features model the offline activities in such paid K-12 online services. For examples, students and their parents receive follow-ups based on their previous class performance and give their satisfaction feedbacks. Another example is that students may request changes to their course schedules.
\end{itemize}

\subsubsection{Time-variant Features}
\label{sec:time_variant}

Besides in-class and out-of-class features, we manually design time-variant features to model the changes of likelihood of students' dropout intentions. Cases like a student just had a class compared to a student had a class two weeks ago should be explicitly distinguished when constructing features. Therefore, we create time-variant features by utilizing a lookback window approach on students' observation sequences. More specifically, for a given timestamp, we only focus on previously observed activities of each student within a period of time. The length of lookback windows varies from 1 to 30 days. Sufficient statistics are extracted as time-variant features from each lookback window. Meanwhile, we compute historical performance features to reflect the teaching experience and performance for each individual teacher. Table \ref{tab:example} shows some examples of time-variant feature we use in our dropout prediction framework.

\begin{itemize}
    \item \textbf{Lookback window features}: The lookback window features aggregate important statistics from students' observations within a fixed-length lookback window, such as the numbers of courses taken in past one, two, three weeks.

    \item \textbf{Historical performance features}: The historical features aggregate each teacher's past teaching performance, which represent the overall teaching quality profiles. They include total numbers of courses and students taught, historical dropout rates, etc.
\end{itemize}

\subsection{Data Augmentation}
\label{sec:label}
According to Assumptions \ref{as:recency} and \ref{as:multistep} and the problem formulation in Section \ref{sec:formulation}, a real-world early warning system is supposed to flexibly support multi-step ahead predictions for each student, i.e., given any future time span $\Delta$, the system computes the probability of student's status $\hat{y}^s_{t^s_{n_s+\Delta}}$. The predicted probability should be able to dynamically adapt when the values of $\Delta$ get changed. The multi-step ahead assumption essentially requires the approach to make predictions at a more fine-grained granularity of <student, timestamp> pair, i.e, $<s, t^s_{n_s+\Delta}>$, instead of student level, i.e., $s$. This poses a challenging question: due to the fact that only about 20\% of K-12 students drop their online classes, \emph{how do we tackle the class imbalance problem  when extracting <student, timestamp> training pairs from a collection of multimodal observation sequences (either completion or dropout) in K-12 online scenarios?}

Let $\mathbf{S}_1$ and $\mathbf{S}_2$ be the set of student indices of dropout and non-dropout students, i.e., $\mathbf{S}_1 = \{i | y^i_{n_i} = dropout \}$, and $\mathbf{S}_2 = \{j | y^j_{n_j} = completion \}$. Let $\mathbf{P}$ and $\mathbf{N}$ be the sets of positive (dropout) and negative (non-dropout) <student, timestamp> pairs. By definition, $\mathbf{P}$ and $\mathbf{N}$ are constructed as follows:

\begin{align}
\mathbf{P} & = \{  <\mathbf{x}^i_{n_i},t^i_{n_i}>  | i \in  \mathbf{S}_1 \} \nonumber \\ 
\mathbf{N} & = \{  <\mathbf{x}^i_k,t^i_k>  | i \in  \mathbf{S}_1, k \in \mathbf{T}_i \backslash t^i_{n_i} \} \nonumber \\
 & \cup \{  <\mathbf{x}^j_k,t^j_k>  | j \in  \mathbf{S}_2, k \in \mathbf{T}_j \} 
 \label{eq:orignial_label}
\end{align}

Similar to many researches such as fraud detection \cite{wei2013effective}, the sizes of $\mathbf{P}$ and $\mathbf{N}$ are typically very imbalanced. While in some cases the class imbalance problem may be alleviated by applying an over-sampling algorithm on the minority class sample set, the diversity of the available instances is often limited. Therefore, in this work, we propose a time-aware data augmentation technique that artificially generates pseudo positive (dropout) <student, timestamp> pairs. 

More specifically, for each dropout student $i$ in $\mathbf{S}_1$, we set a lookback window with length $\Lambda$ where $\Lambda \leq t^i_{n_i} - t^i_{n_{i-1}}$. For each timestamp $\tilde{t}^i_l$ in the lookback window such that

\begin{equation}
    \max(t^i_{n_{i-1}}, t^i_{n_i} - \Lambda) <  \tilde{t}^i_l < t^i_{n_i}.
\end{equation}

\begin{figure}[!tpbh]
    \centering
    \includegraphics[width=\linewidth] {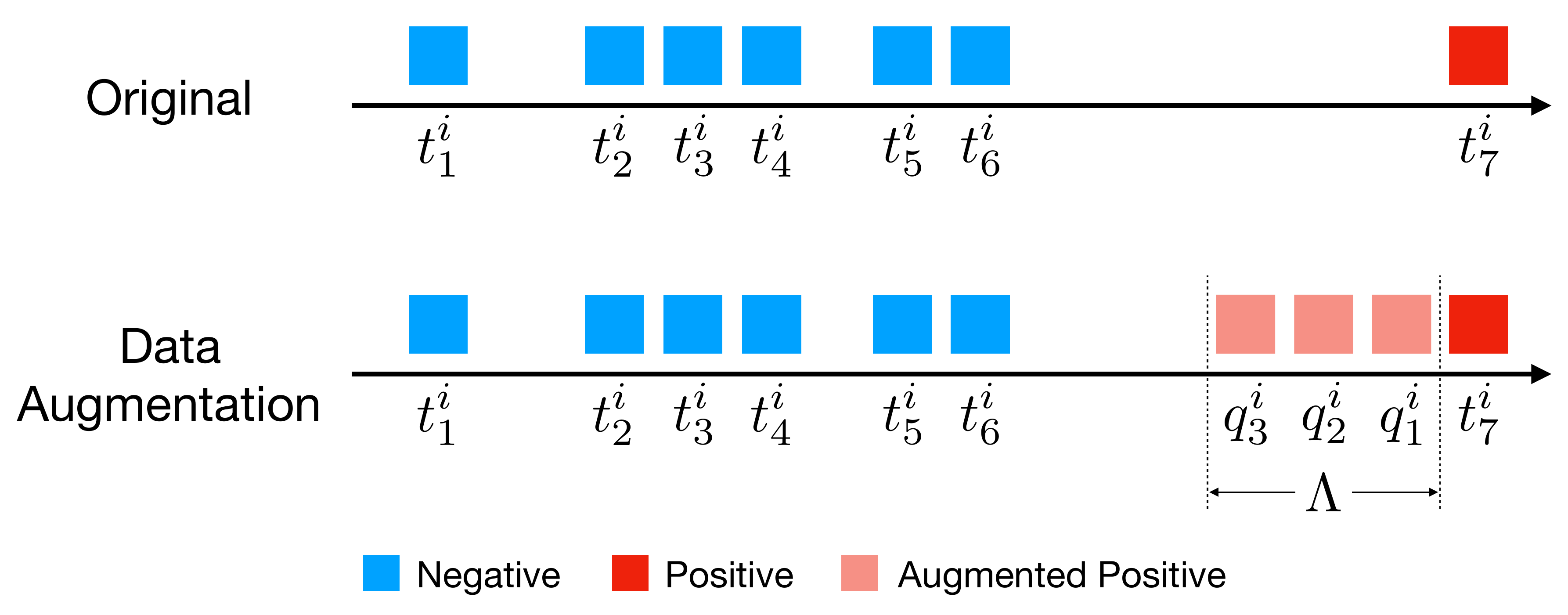}
    \caption{Graphical illustration of the data augmentation technique.}
    \label{fig:augmentation}
\end{figure}

We generate its corresponding pseudo positive training pair $<\tilde{\mathbf{x}}^i_{l}, \tilde{t}^i_l>$ as follows: $\tilde{\mathbf{x}}^i_{l} = \mathcal{F}(\mathbf{X}^s, \mathbf{T}^s)$ where $\mathcal{F}(\cdot, \cdot)$ is the generation function. The choices of $\mathcal{F}(\cdot, \cdot)$ are flexible and vary among different types of features (See Section \ref{sec:features}). In this work, for in-class and out-of-class features, we aggregate all the available features till $\tilde{t}^i_l$ and re-compute the time-variant features according to timestamp $\tilde{t}^i_l$. We use $\tilde{\mathbf{P}}$ to represent the collection of all positive training pairs generated from dropout students in $\mathbf{S}_1$. Figure \ref{fig:augmentation} illustrates how the pseudo positive training pairs are generated.

\begin{figure}[!tpbh]
    \centering
    \includegraphics[width=\linewidth] {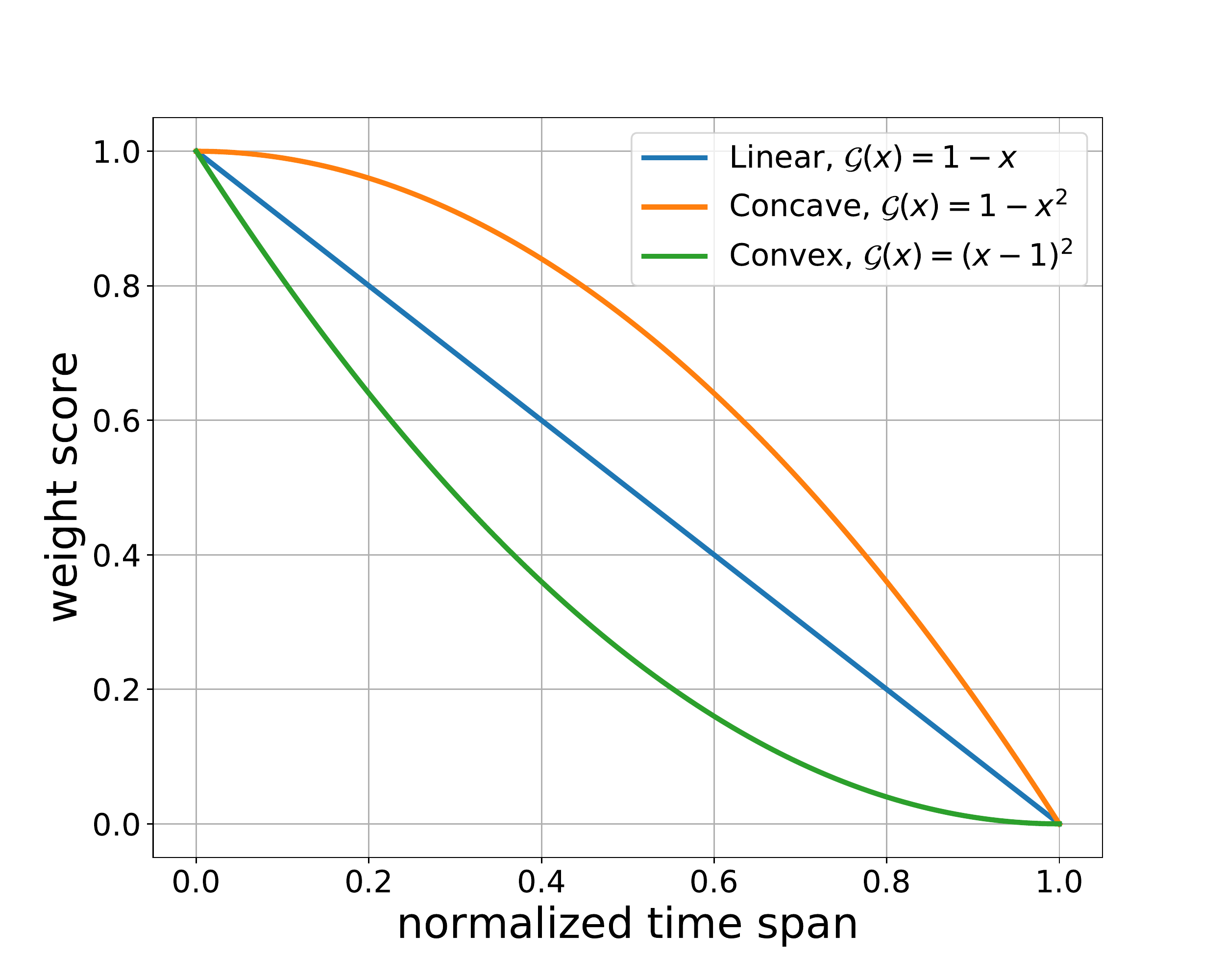}
    \caption{Graphical illustration of different weighting function of $\mathcal{G}(\cdot)$.}
    \label{fig:decay_plot}
\end{figure}

Besides, we assign a time-aware weight to each pseudo positive training pair to reflect the recency effect in Assumption \ref{as:recency}. For each pseudo pair $<\tilde{\mathbf{x}}^i_{l}, \tilde{t}^i_l>$, the corresponding weight $w^i_l$ is computed by

\begin{equation}
\label{eq:weight}
w^i_l = \mathcal{G}(\frac{t^i_{n_i} - \tilde{t}^i_l}{\Lambda})
\end{equation}

\noindent where the weighting function $\mathcal{G}(\cdot)$ takes the normalized time span between each timestamp of pseudo pair and the exact dropout date as input and outputs a normalized weighting score to reflect our confidence on the ``positiveness'' of the simulated training pairs. The closer to the dropout date, the larger the confidence weights should be. The choices of $\mathcal{G}(\cdot)$ are open to any function that gives response values ranging from 0 to 1, such as linear, convex or concave functions illustrated in Figure \ref{fig:decay_plot}.

The effect of different choices of weighting function is discussed in Section \ref{sec:impact_weighting}. The augmented training set $\tilde{\mathbf{P}}$ and the corresponding time-aware weights are used in the model training in Section \ref{sec:learning}.

\subsection{Model Learning}
\label{sec:learning}

In the learning stage, we combine the original training set ($\mathbf{P}$ and $\mathbf{N}$) with the augmented set $\tilde{\mathbf{P}}$ for model training. Even though the data augmentation alleviates the class imbalance problem, i.e., improving the positive example ratio from 0.1\% to 10\%, the imbalance problem still exists. Therefore, we employ the classical weighted over-sampling algorithm on positive pairs to further reduce the imbalance effect. Here, the weights of the original positive examples in $\mathbf{P}$ are set to 1 and pseudo positive examples' weights are computed by $\mathcal{G}(\cdot)$ in Section \ref{sec:label}. Here, since the dropout datasets are usually small compared to other Internet scaled datasets, we choose to use Gradient Boosting Decision Tree\footnote{https://scikit-learn.org/stable/modules/ensemble.html\#gradient-tree-boosting} (GBDT) \cite{friedman2002stochastic} as our prediction model. The GBDT exhibits its robust predictive performance in many well studied problems \cite{ke2017lightgbm,son2015tracking}.

\subsection{Summary}
\label{sec:summary}

The overall model learning procedure of our K-12 online dropout prediction can be summarized in Algorithm \ref{alg:summary}.

\begin{algorithm}[!htbp]
\small \caption{Model learning procedure of the K-12 online dropout prediction.}
\label{alg:summary}
{INPUT:}
\begin{compactitem}
\item A set of K-12 students $\mathbf{S}$ and their corresponding multimodal classroom recordings and activities logs.
\item The length of lookback window $\Lambda$.
\item The choice of weighting function $\mathcal{G}(\cdot)$.
\end{compactitem}
{PROCEDURE:}
\begin{algorithmic}[1]
\STATE // Feature extraction 
\STATE Extract in-class features from multimodal recordings, see Section \ref{sec:inclass}.
\STATE Extract out-of-class features from student activities logs, see Section \ref{sec:post}.
\STATE Extract time-variant features, see Section \ref{sec:time_variant}.
\STATE Concatenate three types of features above. 
\STATE // Label generation and augmentation
\STATE Create original positive and negative training pair sets, i.e., $\mathbf{P}$ and $\mathbf{N}$, see eq.(\ref{eq:orignial_label}).
\STATE Generate the augmented pseudo positive training sets, i.e., $\tilde{\mathbf{P}}$ and the corresponding weights, see eq.(\ref{eq:weight}).
\STATE // Model learning
\STATE Conduct weighted over-sampling on the union of $\mathbf{P}$ and $\tilde{\mathbf{P}}$.
\STATE Train the GBDT model on the over-sampled positive examples and original negative examples.
\end{algorithmic}
{OUTPUT:}
\begin{compactitem}
\item The GBDT dropout prediction model $\Omega$.
\end{compactitem}
\renewcommand*\arraystretch{1.0}
\end{algorithm}

\section{Experimental Evaluation}
\label{sec:exp}
In this section, we will (1) introduce our dataset that is collected from a real-world K-12 online learning platform and the details of our experimental settings (Section \ref{sec:setting}); (2) show that our approach is able to improve the predictive performance when compared to a wide range of classic baselines (Section \ref{sec:performance}); (3) evaluate the impacts of different sizes of lookback windows, different weighting functions in data augmentation and feature combinations (Section \ref{sec:impact_lookback}, Section \ref{sec:impact_weighting} and Section \ref{sec:impact_feature}); and (4) deploy our model into the real production system to demonstrate its effectiveness (Section \ref{sec:online}).

We would also like to note that hyper parameters used in our methods are selected (in all experiments) by the internal cross validation approach while optimizing models’ predictive performances. In the following experiment, we set the size of lookback window to 7 and the impact of window size is discussed in section \ref{sec:impact_weighting}. We choose to use the convex weighting function when conducting pseudo positive data augmentation.

\subsection{Experimental Setting}
\label{sec:setting}

\subsubsection{Data}

To evaluate the effectiveness of our proposed framework, we conduct several experiments on a real-world K-12 online course dataset from a third-party online education platform. We select 3922 registered middle school and high school students from August 2018 and February 2019 as our samples. All the features listed in Section \ref{sec:features} are computed and extracted from daily activity logs on the platform. In our dataset, 634 students choose to drop the class and the dropout rate is 16.16\%. The average time span of the students on the platform is about 86 days, which provide us 338428 observational <student, time stamp> sample pairs in total. We randomly select 80\% of students and use their corresponding <student, time stamp> sample pairs as training set and the remaining 20\% of students' sample pairs for testing propose. The data augmentation technique discussed in Section \ref{sec:label} is only applied in training set.

\subsubsection{Multi-step Ahead Prediction Setting}

To fully examine the dropout prediction performance, we evaluate the model's predictions in terms of different multi-step ahead time spans, i.e, given a current timestamp, we predict the outcome (dropout or non-dropout) in the next $X$ days, where $X$ ranges from $1, 2, \cdots, 14$. 

\subsubsection{Evaluation Metric}

Similar to \cite{gitinabard2018your,tang2018time,gardnermodeling,fei2015temporal,taylor2014likely,he2015identifying}, we evaluate and compare the performance of the different methods by using the Area Under Curve (AUC) score, which is the area under the Receive Operating Characteristic curve (ROC) \cite{fawcett2006introduction}. An ROC curve is a graphic plot created by plotting the true positive rate (TPR) against the false positive rate (FPR). In our dropout prediction scenario, the TPR is the fraction of the ``at-risk'' predicted students who truly drop out. The FPR is the ratio of the falsely predicted ``dropout'' students to the true ones. The AUC score is invariant to data imbalance issue and it does not require additional parameters, for models comparisons. AUC score reaches its best value at 1 and the worst at value 0.

\subsubsection{Baselines}

We compare our proposed approach with the following representative baseline methods: (1) Logistic Regression (LR) \cite{kleinbaum2002logistic}, (2) Decision Tree (DT) \cite{safavian1991survey} and (3) Random Forest (RF) \cite{ho1995random}. LR, DT and RF are all trained on the same set of features defined in Section \ref{sec:features} with our proposed method. The training set is created by using eq.(\ref{eq:orignial_label}).

\subsection{The Overall Prediction Performance}
\label{sec:performance}

The results of these models are shown in Figure \ref{fig:overall}. As we can see from the Figure  \ref{fig:overall}, we have the following observations:

\begin{figure}[!tpbh]
    \centering
    \includegraphics[width=\linewidth]{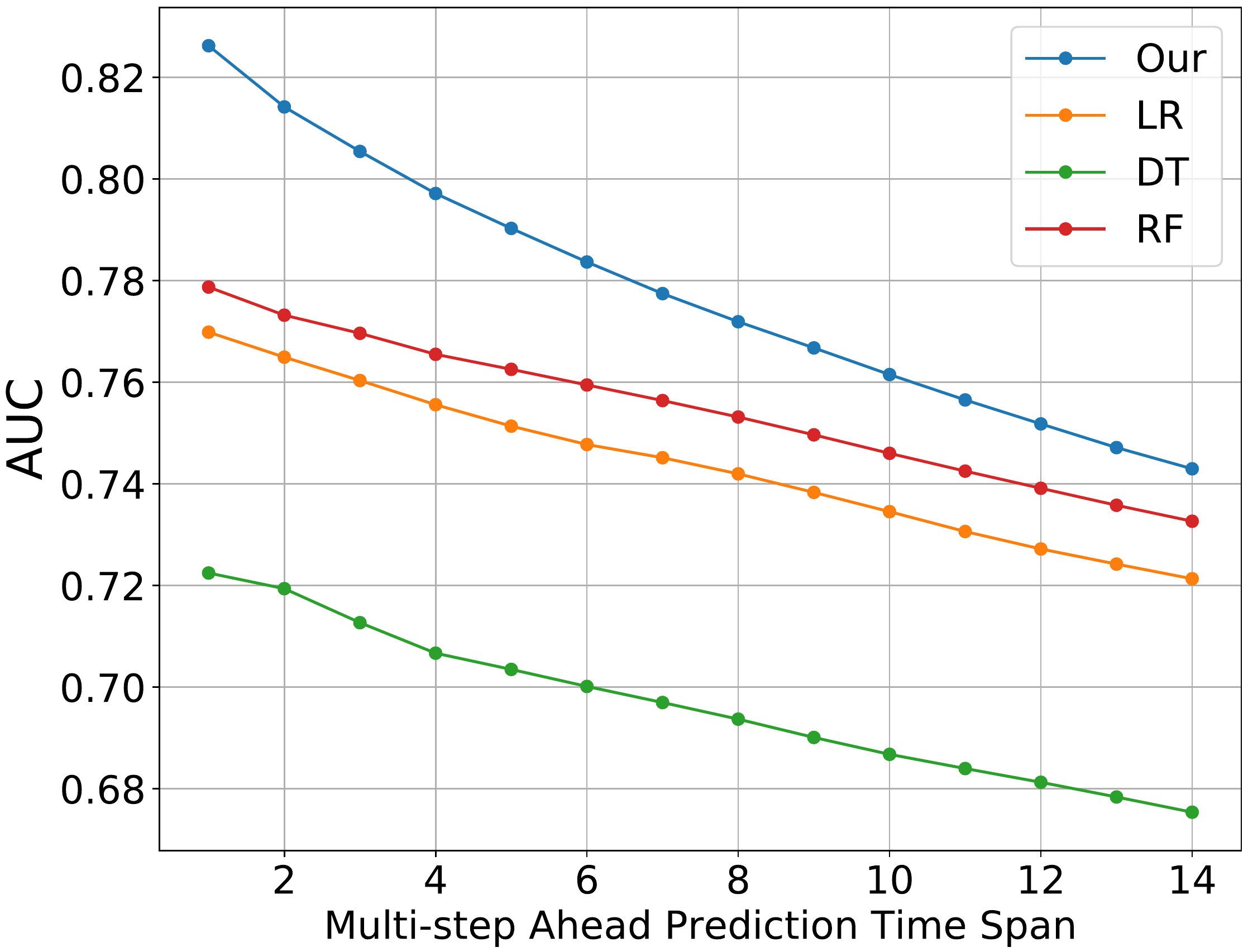}
    \caption{The overall prediction performance with different multi-step ahead time spans in terms of AUC scores.}
    \label{fig:overall}
\end{figure}

\begin{itemize}
\item First, our model outperforms all other methods in terms of AUC scores on different future time spans, which demonstrates the effectiveness of our approaches with positive data augmentation. By adding more diverse pseudo positive training pairs with the corresponding decaying confidence weights, the GBDT model is able to learn the dropout patterns from multiple factors. 

\item Second, as we increase the lengths of time spans of multi-step ahead prediction, all the models' performances decrease accordingly. Our approach achieves AUC score of 0.8262  in the task of next day prediction while the performance downgrades to 0.7430 in the next two-week prediction task. We believe this is because of the truth negative mistakes the models make, i.e., the model thinks the students will continue but they drop classes in next two weeks. This indicates that without knowing more information from the students, the ML models have very limited ability in predicting the long-term outcomes of student status, which also reflects the fact that there are many factors that could lead to the dropouts.

\item Third, comparing LR, DT, and RF, we can see, the DT achieves the worst performance. This is because of its instability. With small number of training data, the DT approach suffers from fractional data turbulence. The RF approach remedies such shortcomings by replacing a single decision tree with a random forest of decision trees and the performance is boosted. Meanwhile, as a linear model, the LR is not powerful enough to accurately capture the dropout cases.
\end{itemize}

\subsection{Impact of Sizes of Lookback Windows}
\label{sec:impact_lookback}

\begin{figure}[!tpbh]
    \centering
    \includegraphics[width=\linewidth] {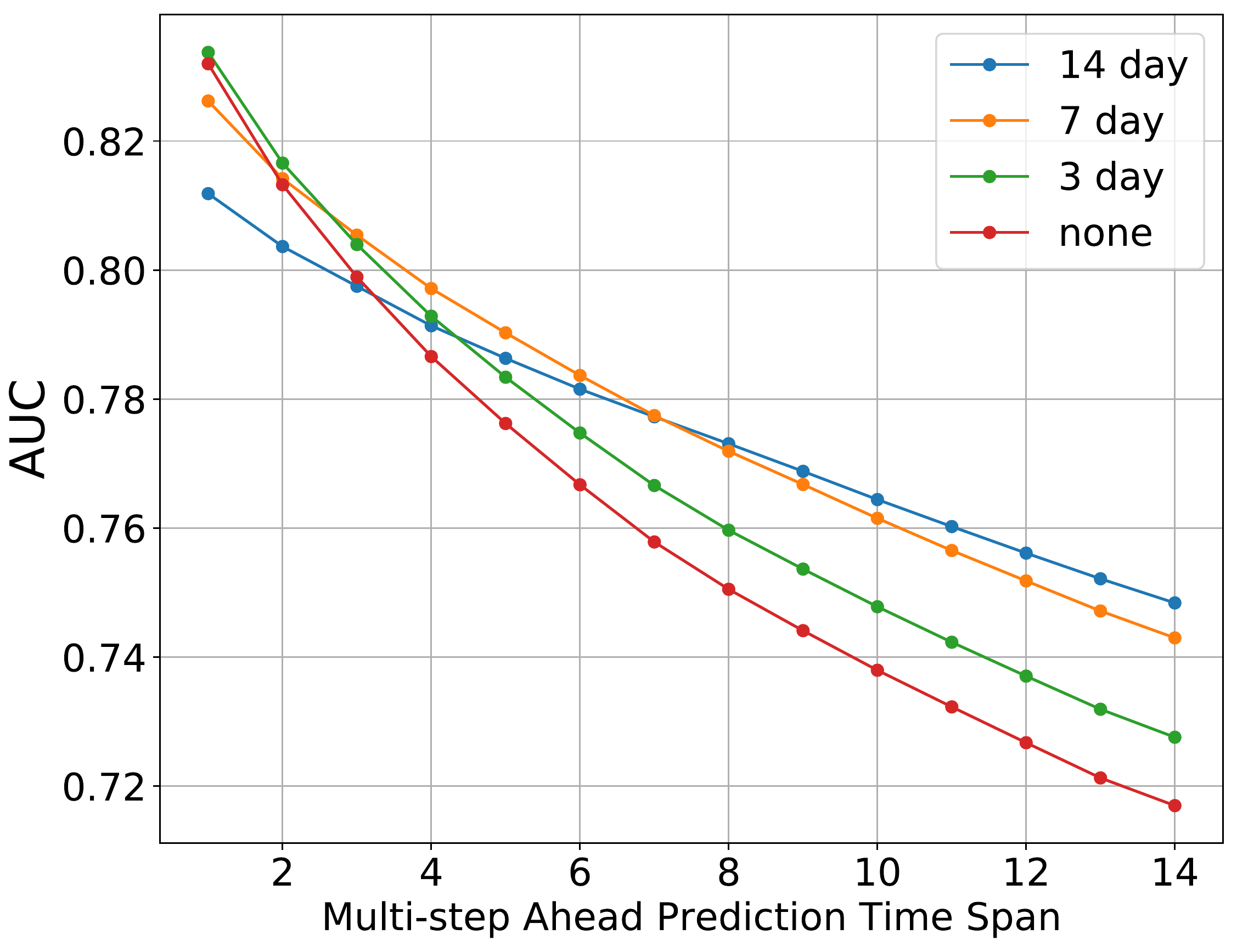}
    \caption{Models trained on data augmented by different size of lookback windows with different multi-step ahead time spans in terms of AUC scores. \emph{none} represents the model training without any lookback data augmentation.}
    \label{fig:lookback}
\end{figure}

As we can see, the number of augmented positive training pairs is directly determined by the size of lookback window $\Lambda$. Therefore, to comprehensively understand the performance of our proposed approach, we conduct experimental comparisons on different sizes of lookback windows. We vary the window size from 3, 7, and 14. Meanwhile, we add a baseline with no data augmentation. The results are shown in Figure \ref{fig:lookback}.

From Figure \ref{fig:lookback}, we can see that the size of lookback windows has a positive relationship on AUC scores with the length of time span in multi-step ahead prediction. When conducting short-term dropout predictions, models trained on data augmentation with smaller size of lookback window outperform others. As we gradually increase the time span of future predictions, the more the model looks back, the higher the prediction AUC score it achieves. Overall, the model trained with 7-day lookback window has the best performance across different multi-step ahead time spans in terms of AUC scores.

\subsection{Impact of Different Weighting Functions}
\label{sec:impact_weighting}

\begin{figure}[!tpbh]
    \centering
    \includegraphics[width=\linewidth] {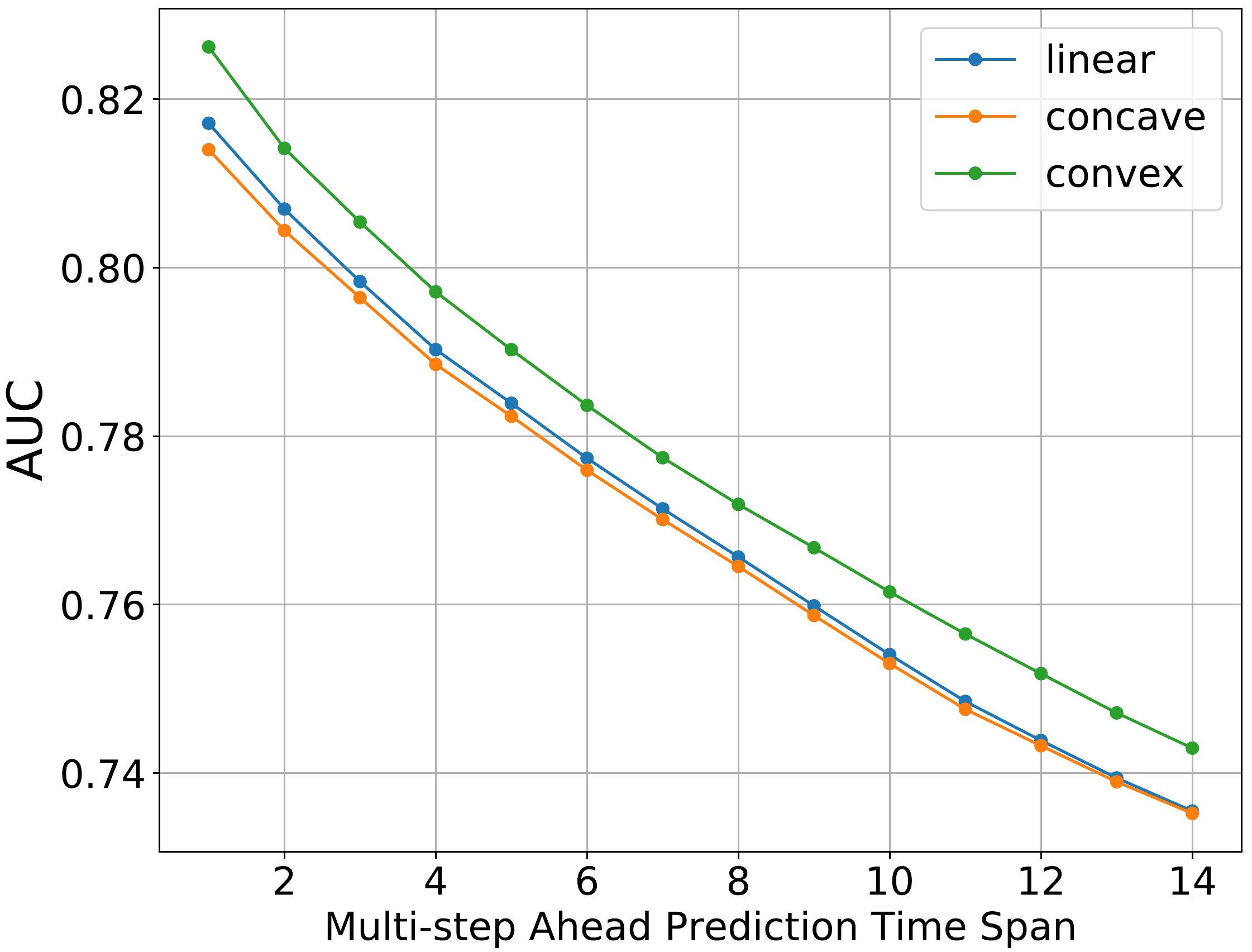}
    \caption{Models trained on data augmented by different choices of weighting functions with different multi-step ahead time spans in terms of AUC scores.}
    \label{fig:weight_function}
\end{figure}

In this section, we examine the performance changes by varying the forms of weighting functions. More specifically, we compare the prediction results of using the convex function to results of the other choices. The results are shown in Figure \ref{fig:weight_function}. As we can see from Figure \ref{fig:weight_function}, the convex option outperforms other choices by a large margin across all different multi-step ahead time spans. When computing the over-sampling weights of pseudo training examples, the convex function gives more weights to the most recent examples, i.e., examples close to the timestamp of true dropout observations. This also confirms the necessity of considering the recency effect assumption (Assumption \ref{as:recency}) when building the dropout prediction framework.

\subsection{Impact of Different Features}
\label{sec:impact_feature}

\begin{table*}[!tpbh]
    \centering
    \caption{Experimental results of different types of features and different lengths of multi-step ahead time span in terms of AUC scores.}
    \label{tab:feature_prediction}
    \begin{tabular}{l|c|c|c|c|c|c|c}
    \toprule
     &  In &  Out  &  Time  &  In+Time & Out+Time & In+Out & In+Out+Time\\
    \midrule
    Multi-step ahead time span - 7 day & 0.6249 & 0.7764 & 0.6992 & 0.7145 & 0.7759 & 0.7768 & \bf{0.7774} \\
    Multi-step ahead time span - 14 day & 0.6251 & 0.7386 & 0.6766 & 0.6932 & 0.7393 & 0.7420 & \bf{0.7430} \\
    \bottomrule
    \end{tabular}
\end{table*}

In this subsection, we systematically examine the effect of different types of features by constructing following model variants:

\begin{itemize}
\item In: only the in-class features are used.
\item Out: only the out-of-class features are used.
\item Time: only the time-variant features are used.
\item In+Time: it eliminates the contribution of \emph{Out} features and only uses features from \emph{In} and \emph{Time}.
\item Out+Time: it eliminates the contribution of \emph{In} features and only uses features from \emph{Out} and \emph{Time}.
\item In+Out: it eliminates the contribution of \emph{Time} features and only uses features from \emph{In} and \emph{Out}.
\item In+Out+Time: it uses the combination of all the features from \emph{In}, \emph{Out} and \emph{Time}.
\end{itemize}

Meanwhile, we also consider different multi-step ahead prediction settings, i.e., next 7-day prediction and next 14-day prediction and the prediction results are shown in Table \ref{tab:feature_prediction}. From Table \ref{tab:feature_prediction}, we observe that (1) by considering all three types of features individually, the model trained from \emph{Out} features yields the best performance. Moreover, when comparing \emph{In}, \emph{Out} to \emph{In+Time}, \emph{Out+Time}, we obtain the significant performance improvement by adding \emph{Out} features. These indicate the fact that dropout prediction for K-12 online scenarios are very different from MOOC based dropout prediction. The out-of-class activities and the quality of the service play an extremely important role in the prediction task; and (2) by utilizing all the sets of features, we could be able to achieve the best results in both prediction tasks.

\subsection{Online Performance}
\label{sec:online}

We deployed our at-risk student warning system in the real production environment on a third-party platform between February 2nd, 2019 to April 1st, 2019. To watch the system performance in practice, we conduct the next-day prediction task where the system predicts the dropout probability for each on-going student at 6 am in the morning. All the students are ranked by their dropout probabilities and the top 30\% of students with highest probabilities are marked as at-risk students. At the end of each day, we obtain the real outcome of all the students who drop the class. We conduct the overlap comparison between the predicted top at-risk students (30\% of total students) and the daily dropouts and we are able to achieve that more than 70\% of dropout students are detected by the system.

\section{Conclusion}
\label{sec:conclusion}
In this paper, we present an effective at-risk student identification framework for K-12 online classes. Compared to the existing dropout prediction researches, our approach considers and focuses on the challenging factors such as multiple modalities, length variability, time sensitivity, class imbalance problems when learning from real-world K-12 educational data. Our offline experimental results show that our approach outperforms other state-of-the-art prediction approaches in terms of AUC scores. Furthermore, we deploy our model into a production environment and we are able to achieve that more than 70\% of dropout students are detected by the system. In the future, we plan to explore the opportunity of using deep neural networks to capture heterogeneous information in the K-12 online scenarios to enhance the existing prediction pipeline.

\section{Acknowledgments}
\label{sec:acknowledgements}
The authors would like to thank Nan Jiang from Peking University and Lyu Qing from Tsinghua University, for their contributions to this research during their internship in TAL Education Group.

%
\bibliographystyle{abbrv}
\bibliography{edm2020}  
\balancecolumns
\end{document}